\documentclass[a4paper,fleqn,numbers]{cas-dc}

\usepackage[numbers]{natbib}
\usepackage{graphicx}
\usepackage{amsmath}
\usepackage{amssymb}
\usepackage{soul}
\usepackage{xcolor}
\usepackage{booktabs}
\usepackage{lineno}
\usepackage{algorithm}
\usepackage{algorithmicx}
\usepackage{algpseudocode}
\usepackage{makecell}
\usepackage{empheq}
\usepackage{subcaption}
\usepackage{balance}

\begin{document}
\let\WriteBookmarks\relax
\def\floatpagepagefraction{1}
\def\textpagefraction{.001}
\shorttitle{Control of hybrid wind-wave energy systems using reinforcement learning}
\shortauthors{Z. Lin et~al.}

\title [mode = title]{Control of hybrid wind-wave energy systems using reinforcement learning}                      
\tnotemark[1]

\tnotetext[1]{This wors was supported by the Sustainable Energy Authority of Ireland through the RDD Energise Fellowship under Grant No. 25/RDDF/793.}

\author[1]{Zechuan Lin}[]
\cormark[1]
\ead{Zechuan.Lin@mu.ie}

\author[2]{Kemeng Chen}[]
\ead{ckm22@mails.tsinghua.edu.cn}

\author[2]{Maosen Fan}[]
\ead{fms24@mails.tsinghua.edu.cn}

\author[3,4]{Xiaofan Li}[]
\ead{lixf@hku.hk}

\author[2]{Xi Xiao}[]
\ead{xiao_xi@tsinghua.edu.cn}

\author[1]{John V. Ringwood}[]
\ead{John.Ringwood@mu.ie}

\affiliation[1]{organization={Centre for Ocean Energy Research, Maynooth University},
                city={Maynooth},
                postcode={W23 F2H6}, 
                state={Co. Kildare},
                country={Ireland}}
\affiliation[2]{organization={Department of Electrical Engineering, Tsinghua University},
                city={Beijing},
                postcode={100084}, 
                country={China}}
\affiliation[3]{organization={Department of Mechanical Engineering, The University of Hong Kong},
                city={Hong Kong SAR},
                country={China}}
\affiliation[4]{organization={Swire Institute of Marine Science, The University of Hong Kong},
                city={Hong Kong SAR},
                country={China}}
                
\cortext[cor1]{Corresponding author}

\begin{abstract}
Integrating wave energy converters (WECs) with floating offshore wind turbines (FOWTs), to form hybrid wind-wave energy (HWWE) systems, is a promising approach to achieve further cost reduction for offshore renewable energy. In such systems, the control of the integrated WECs plays an important role, with the potential to generate additional wave energy while simultaneously suppressing floating platform motion. However, HWWE systems are characterized by complex dynamics, making accurate modelling only viable through numerical simulation, and posing significant challenges for control design. This paper proposes a reinforcement learning (RL) control framework for HWWE systems, in which the real-time control policy is learned directly through interactions with high-fidelity simulation. A numerical model is established for a HWWE system consisting of an IEA 15 MW wind turbine, a VolturnUS semi-submersible platform, and three torus-type WECs, which is then employed as the RL training environment. Control performance is evaluated in terms of both wave energy generation and platform motion reduction, two competing objectives, from a Pareto perspective. It is shown that the proposed RL controller achieves substantial Pareto improvements over conventional control strategies, e.g., over 75\% higher wave energy capture at the same platform motion level, or nearly 50\% lower motion at the same energy capture level, thereby significantly extending the attainable performance boundary of HWWE systems.
\end{abstract}

\if false
\begin{graphicalabstract}
\includegraphics{figs/cas-grabs.pdf}
\end{graphicalabstract}
\fi

\if false
\begin{highlights}
\item Reinforcement learning control developed for hybrid wind-wave energy systems.
\item Control objectives include wave energy maximization and platform motion suppression. 
\item Wave energy--platform stability tradeoff evaluated from a Pareto perspective.
\item Homogeneous and heterogeneous reactive control taken as benchmarks. 
\item Reinforcement learning substantially advances Pareto front over reactive control.
\end{highlights}
\fi


\maketitle

\section{Introduction}
Large-scale development of offshore renewable energy could play a pivotal role in accelerating the global energy transition and achieving net-zero emissions. In recent years, the concept of integrating wave energy converters (WECs) with floating offshore wind turbines (FOWTs), to form hybrid wind-wave energy (HWWE) systems, has received increasing attention \cite{wan2024review, hallak2025overview}. HWWE systems offer a number of advantages over FOWT-only or WEC-only systems, including a shared floating platform and mooring system with lower per-unit electricity cost \cite{hallak2025overview}, more efficient exploitation of ocean space, and motion suppression effects provided by WEC integration \cite{yi2026study}.

A number of HWWE system concepts have been proposed, such as the spar-torus combination \cite{muliawan2013dynamic} and the combination of semi-submersible FOWTs with point-absorber WECs \cite{han2024dynamic}, flap-type WECs \cite{gao2016comparative}, or oscillating water columns (OWCs) \cite{zhang2022coupled}. In general, HWWE systems are characterized by strongly coupled multibody, multi-degree-of-freedom, and multiphysics dynamics. While relatively mature tools exist for modelling standalone FOWTs and WECs, no established commercial software is currently available for coupled FOWT-WEC analysis. Consequently, a variety of numerical frameworks, mainly aiming to couple individual FOWT and WEC simulation tools, have been developed \cite{celesti2025towards}, such as SIMO-RIFLEX-AeroDyn \cite{gao2016comparative}, OpenFAST-AQWA \cite{han2024dynamic}, OpenFAST-WEC-Sim \cite{wang2025numerical}, OpenFAST-WAFDUT with an in-house multibody dynamic solver \cite{zhou2023coupled}, computational fluid dynamics (CFD) simulation \cite{yi2026study}, as well as analytical models in the frequency domain \cite{zhu2024analytical}. Meanwhile, valuable wave tank experiments have been performed for model validation and design analysis \cite{gao2016comparative, michailides2016experimental, gaspar2021compensation}. Overall, numerical and experimental studies have demonstrated the advantages of HWWE systems over their wind-only counterparts, including increased energy capture and reduced platform pitch motion \cite{sergiienko2025statistical}. The performance of HWWE systems is further improved through design optimization of WEC buoy shape, size, layout, etc \cite{wang2022wec, neshat2024enhancing, hu2020optimal, cao2023wecs}.

Throughout existing research, it has been recognized that the control of WECs, through power take-off (PTO) systems, has an important impact on energy capture and platform motion response \cite{si2021influence}. However, while control of standalone WECs has been extensively studied, WEC control in HWWE systems remains a less explored area \cite{celesti2025towards}. Most existing studies adopt passive PTO control, usually described by a linear damping coefficient; in addition, constant or quadratic damping are also considered \cite{zhu2024analytical, wang2026enhancing}. In some studies, the damping coefficient is selected according to the energy-maximizing (impedance-matching) condition of standalone WECs \cite{cao2023wecs, bayat2026multidisciplinary}. The impact of damping is further examined in \cite{han2024dynamic}, and various algorithms are employed to optimize the damping coefficient, taking into account both the energy and stability objectives \cite{wang2025numerical, wang2026enhancing}. Reactive control, in which a PTO stiffness term is added to the damping force, is further covered by a number of studies. A small positive stiffness is used, in \cite{li2022power}, to provide a mechanical restoring effect. The effect of PTO stiffness is thoroughly examined in \cite{hu2020optimal, si2021influence}, where positive stiffness is shown to reduce wave energy capture while improving platform stability, whereas lower or negative stiffness tends to produce the opposite effect.

Other relatively simple control strategies are also developed. A gain-scheduling strategy is proposed for damping control of integrated OWCs in \cite{zhang2022coupled}. Also, a bang-bang control law is employed to switch the damping coefficient according to platform motion states \cite{chen2022load}. A wave-period-dependent switching strategy is proposed to control OWC valves, following an analysis of control influence on the response amplitude operator \cite{aboutalebi2023control}. A `sky-hook' control strategy is developed for the integrated OWCs, and a linearized model is developed to analyze the impact of the control coefficient through root-locus analysis \cite{zhu2025experiment}. Fuzzy logic control is investigated in \cite{ahmad2023fuzzy, m2023fuzzy}, where empirical mappings from platform motion to desired OWC valve commands are established. 

In general, the above simple, or empirically designed, controllers are straightforward to implement, and can be tuned directly against numerical simulation or experiment. However, such controllers inherently offer limited control flexibility; e.g., reactive control usually contains only two tunable coefficients (damping and stiffness), which can lead to substantial suboptimality under the complex dynamics of HWWE systems, as well as varying wind-wave conditions. 

In contrast, model predictive control (MPC) has been explored for HWWE systems in a few studies. In \cite{zhu2022optimal}, the hybrid platform dynamics are linearized around an equilibrium point to obtain the prediction model, and real-time PTO forces are solved via online optimization. In \cite{zhao2024multi}, MPC is applied to a HWWE system with an M4 WEC, where multiple objectives, including energy capture and platform motion suppression, are taken into account, to achieve a suitable balance. Related works also include MPC for co-located wind-wave energy systems (i.e., without mechanical connection) \cite{meng2023co} and an active pumping system inside a FOWT (without WECs) \cite{wang2026multi}. However, despite the optimization-based formulation, MPC relies on simplified, control-oriented models to enable online optimization, which may introduce substantial model mismatch \cite{zhu2022optimal}, and consequently degrade control performance. 

Hence, a key research gap in HWWE control lies in developing a controller that offers flexible real-time control beyond empirical control laws, while being optimized directly against high-fidelity simulations or experiments \cite{celesti2025towards}. To address this gap, this paper proposes, for the first time, a reinforcement learning (RL) control framework for HWWE systems. RL offers unique advantages, by using neural networks (NNs) to learn a real-time control policy directly through interaction with the simulation environment, and has seen successful application in standalone WECs \cite{chen2024design, chen2026multi}. Moreover, once trained, the NN-based control policy can be evaluated efficiently in real time without solving online optimization problems, making RL a promising solution for the complex control of HWWE systems.

In this paper, a numerical model of a HWWE platform, incorporating coupled aero-hydro-structural-mooring dynamics, is established using MOST \cite{sirigu2022development} and WEC-Sim \cite{shabara2024review}. Linear damping and reactive control are first developed and examined, to serve as control benchmarks. This includes both the homogeneous (HOM) and heterogeneous (HET) reactive control configurations, where all WECs use identical coefficients in the former case, and independently selected coefficients in the latter. Importantly, control performance is evaluated in terms of both wave energy generation and platform motion suppression, two conflicting control objectives, from a Pareto perspective. A multi-objective Bayesian optimization framework is employed to identify the Pareto fronts of HOM and HET reactive control. With these clear benchmarks, an RL control framework is then designed, which adopts the proximal policy optimization (PPO) algorithm \cite{schulman2017proximal}. The RL controller targets a reward function involving energy and stability objectives, which are balanced by a weighting factor, and is trained through iterative interactions against the numerical simulation environment. A main finding of this work is the strong performance demonstrated by RL control, which achieves a substantially improved Pareto front compared with HOM and HET reactive control, simultaneously enabling higher wave energy generation and lower platform motion, and thereby expanding the achievable performance boundary of HWWE systems.

The remainder of this paper is organized as follows. The considered HWWE system and the numerical model are described in Section II. The benchmark HOM and HET reactive controllers are developed and analyzed in detail in Section III. The RL controller, and the associated training framework, are presented in Section IV and comprehensively evaluated in Section V. Conclusions are drawn in Section VI. 

\section{Numerical modelling}
\subsection{System description}
\begin{figure}
    \centering
    \includegraphics[width=0.85\columnwidth]{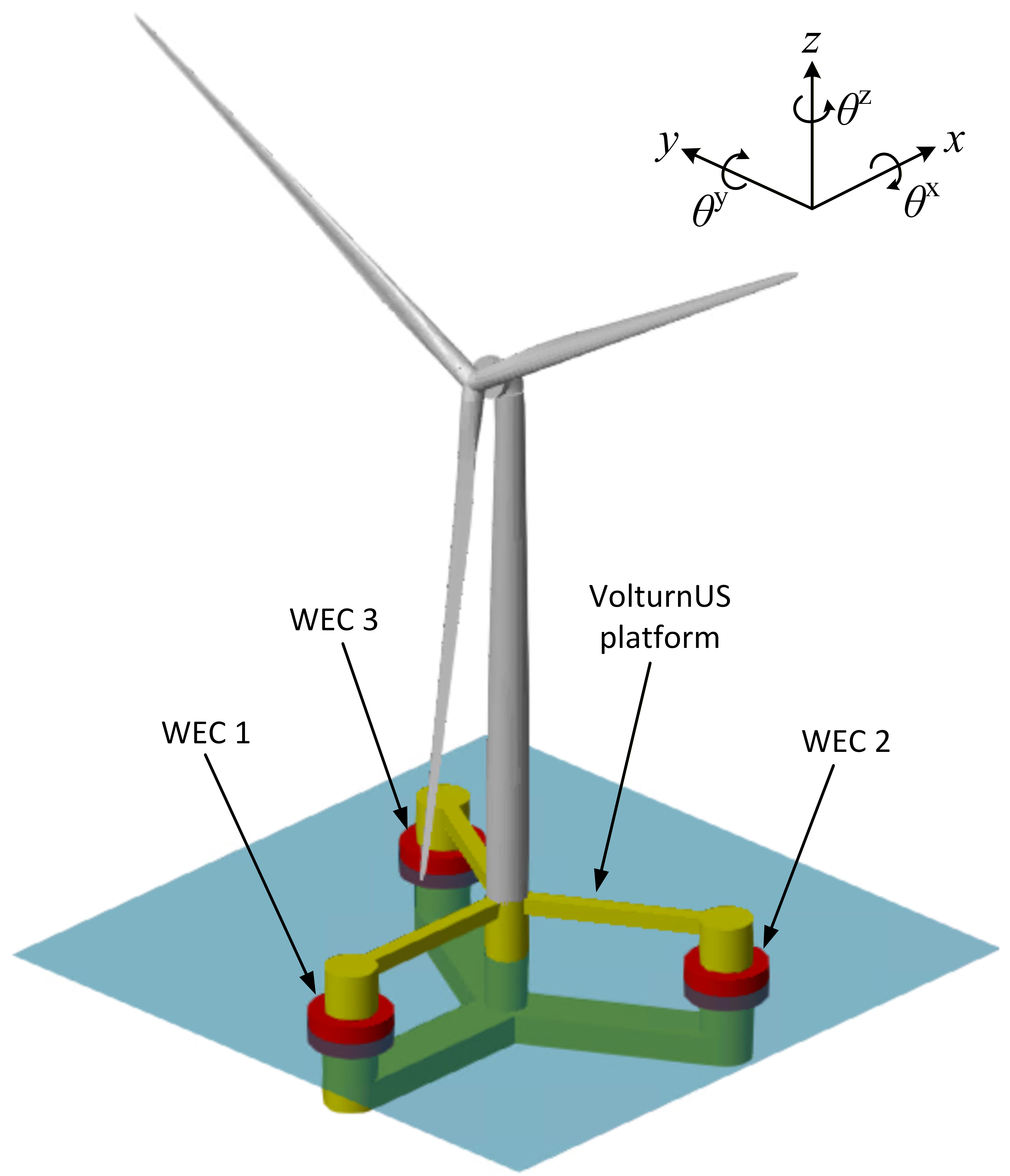}
    \caption{Structure of the HWWE platform. Yellow part: the VolturnUS platform. Red part: the torus-type WECs.}
    \label{fig_platform}
\end{figure}

In this study, a HWWE platform, consisting of an IEA 15 MW wind turbine \cite{gaertner2020definition}, a VolturnUS semi-submersible platform \cite{allen2020definition}, a three-point symmetrical mooring system, and three torus-type cylindrical WECs is considered, as shown in Fig. \ref{fig_platform}; key parameters are listed in Table \ref{tab_platform}. This type of hybrid platform is also investigated in, e.g., \cite{wang2025numerical, neshat2024enhancing, wang2026enhancing}. The integrated WECs move along the outer columns in the vertical direction, i.e., in relative heave motion. The WEC size is selected to be relatively small compared to the dominant wavelength, so that the WECs approach the conditions of point absorbers. Each WEC is connected to the platform through a PTO system, so that electrical power can be generated from relative heave motion. The considered PTO system can be either a linear generator system \cite{chen2026thrust} or a mechanical-transmission-based rotary generator system \cite{huang2024improving}. In both cases, the PTO is able to produce a commanded force on the WEC, and the PTO power can be bidirectional. Accordingly, the control task is to determine the optimal PTO force commands, in order to optimize the overall performance of the HWWE system. 

\begin{table}
\centering
\caption{Main parameters of the HWWE system. For detailed definitions of the wind turbine and VolturnUS platform, see \cite{gaertner2020definition, allen2020definition}.}
\label{tab_platform}
\begin{tabular}{lll}
\hline
Category & Parameter & Value \\
\hline
Platform & Arm length &  51.75 m\\
         & Outer column diameter &  12.5 m\\
         & Inner column diameter &  10 m\\
         & Freeboard &  15 m\\
         & Draft &  20 m\\
\hline
WEC      & Outer diameter &  20 m\\
         & Freeboard & 4 m\\
         & Draft &  4 m\\
\hline
Mooring  & Length      & 850 m \\
         & Diameter    & 0.33 m \\
         & Linear mass & 685 kg/m \\
         & Water depth & 200 m \\
\hline
PTO      & Friction coefficient $R_{\mathrm{fric}}$ & 30 kN/(m/s) \\
         & Maximum force $F_{\max}$ & 2000 kN \\
         & Loss coefficient $\alpha$ & 1.2e-5 kW/(kN)$^2$ \\
\hline
\end{tabular}
\end{table}

\begin{figure}
    \centering
    \begin{subfigure}{1\columnwidth}
        \centering
        \includegraphics[width=\linewidth]{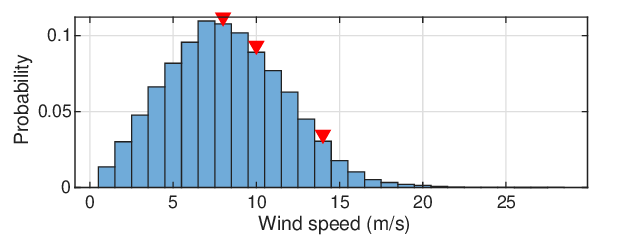}
        \caption{Wind speed distribution.}
        \label{fig_wind_dist}
    \end{subfigure}

    \begin{subfigure}{1\columnwidth}
        \centering
        \includegraphics[width=\linewidth]{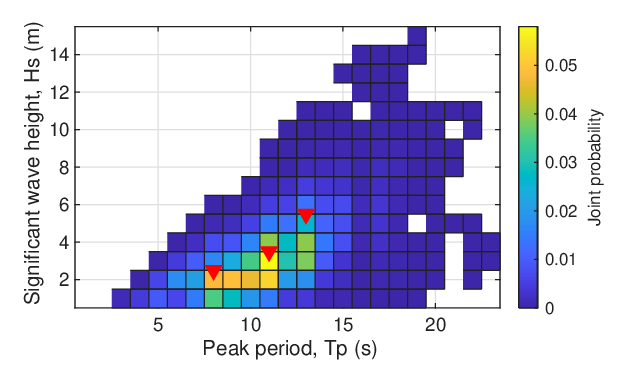}
        \caption{Joint distribution of significant wave height (Hs) and peak period (Tp).}
        \label{fig_wave_dist}
    \end{subfigure}

    \caption{Wind and wave conditions. The selected representative conditions are marked with red triangles.}
    \label{fig_wind_wave_dist}
\end{figure}

The wind and wave conditions used for the case study are obtained from the M4 buoy measurements of the Irish Weather Buoy Network, operated by the Marine Institute, Ireland \cite{marineinstitute2018imdbon}. The buoy is located in the Atlantic Ocean to the northwest of Ireland at 55.0$^\circ$N, 10.0$^\circ$W. The selected data covers the five-year period from 2021 to 2025 and includes hourly measured wind speed (WS), significant wave height (Hs), and peak wave period (Tp). The probability distributions of wind and wave conditions are shown in Fig. \ref{fig_wind_wave_dist}. For ease of testing and illustration, three representative wind-wave conditions, with (WS, Hs, Tp) being (8 m/s, 2 m, 8 s), (10 m/s, 3 m, 11 s), and (14 m/s, 5 m, 13 s), respectively, are selected for the case study. These conditions represent progressively increasing wind-wave energy levels and cover a broad operating range, including the main energy production region, and more energetic conditions that impose greater challenges on platform stability. It is assumed that the mean (turbulent) wind direction and the wave propagation direction are both aligned with the x-axis, with the HWWE platform oriented to face the prevailing incoming wind and waves, as shown in Fig. \ref{fig_platform}.

\subsection{Numerical modelling framework}
The numerical model of the HWWE platform is developed using MOST, a simulation tool developed by Politecnico di Torino \cite{sirigu2022development}. MOST is built upon WEC-Sim \cite{shabara2024review} and extends its capability by adding wind turbine models. Hence, MOST enables coupled simulation of HWWE systems in a unified MATLAB environment, which is particularly convenient for control design. The accuracy of MOST for simulating the VolturnUS platform has been verified against commercial software \cite{sirigu2022development} and through experimental campaigns \cite{niosi2024comparison, yu2025modelling}. This section outlines the fundamental modelling approaches adopted in MOST.

\subsubsection{Multibody dynamics}
The equation of motion of the multibody system can be expressed, in a general form, as
\begin{equation}
\mathbf{M}\left(\boldsymbol{\xi}(t)\right)\ddot{\boldsymbol{\xi}}(t)
+\mathbf{h}\left(\boldsymbol{\xi}(t),\dot{\boldsymbol{\xi}}(t)\right)
=\mathbf{F}(t)+\mathbf{F}_{\mathrm{c}}(t),
\label{eq:constrained_eom}
\end{equation}
where 
$\boldsymbol{\xi}(t)=
\left[\boldsymbol{\xi}_0^{\mathrm{T}}(t),
\boldsymbol{\xi}_1^{\mathrm{T}}(t),
\boldsymbol{\xi}_2^{\mathrm{T}}(t),
\boldsymbol{\xi}_3^{\mathrm{T}}(t)\right]^{\mathrm{T}}$
is the generalized-coordinate vector of the multibody system, with $\boldsymbol{\xi}_i(t)=[x_i(t),y_i(t),z_i(t),\theta^{\mathrm{x}}_i(t),\theta^{\mathrm{y}}_i(t),\theta^{\mathrm{z}}_i(t)]^{\mathrm{T}}$ describing the six-degree-of-freedom pose of the $i$-th body, in terms of surge, sway, heave, roll, pitch, and yaw, respectively, as defined in Fig. \ref{fig_platform}; here, $i=0$ denotes the VolturnUS platform, and $i=1,2,3$ denote the three WECs, respectively. $\mathbf{M}(\boldsymbol{\xi}(t))$ is the mass matrix, $\mathbf{h}\left(\boldsymbol{\xi}(t),\dot{\boldsymbol{\xi}}(t)\right)$ collects the additional inertial terms, including Coriolis, centrifugal, and gyroscopic effects, $\mathbf{F}(t)$ is the load vector resulting from all non-constraint (aerodynamic, hydrodynamic, hydrostatic, mooring, mechanical friction, and PTO) loads, and $\mathbf{F}_{\mathrm{c}}(t)$ is the constraint load vector associated with the reaction forces and moments required to enforce the kinematic constraints between the platform and WECs (WECs are constrained to move only vertically relative to the platform):
\begin{equation}
\mathbf{g}\left(\boldsymbol{\xi}(t)\right)=\mathbf{0},
\label{eq:constraint_general}
\end{equation}
where $\mathbf{g}(\cdot)$ denotes the constraint operator. The constraint load $\mathbf{F}_{\mathrm{c}}(t)$ is related to \eqref{eq:constraint_general} through
\begin{equation}
\mathbf{F}_{\mathrm{c}}(t) = -\left(\frac{\partial \mathbf{g}}{\partial \boldsymbol\xi}\right)^{\mathrm{T}}\boldsymbol{\lambda}(t),
\label{eq:lagrange_multiplier}
\end{equation}
where $\boldsymbol{\lambda}(t)$ is the vector of Lagrangian multipliers. Within MOST, constrained multibody dynamics \eqref{eq:constrained_eom} \eqref{eq:constraint_general} \eqref{eq:lagrange_multiplier} are solved using Simscape Multibody, with a vertical prismatic joint connecting each WEC to the platform. 

\subsubsection{Aerodynamic loads}
In MOST, the aerodynamic loads acting on the turbine blades can be described as
\begin{equation}
\mathbf{F}_{\mathrm{aero}}(t)
=\mathcal{A}
\left(
\mathbf{u}_{\mathrm{w}}(\mathbf{r},t),
\boldsymbol{\xi}_{\mathrm{h}}(t),
\dot{\boldsymbol{\xi}}_{\mathrm{h}}(t),
\psi_{\mathrm{r}}(t),
\dot{\psi}_{\mathrm{r}}(t),
\phi(t)
\right),
\label{eq:aero_force}
\end{equation}
where $\mathbf{u}_{\mathrm{w}}(\mathbf{r},t)$ denotes the wind velocity field, with $\mathbf{r}$ denoting the spatial position, $\boldsymbol{\xi}_{\mathrm{h}}(t)$ and $\dot{\boldsymbol{\xi}}_{\mathrm{h}}(t)$ are the hub pose and the corresponding translational and angular velocities, respectively, $\psi_{\mathrm{r}}(t)$ and $\dot{\psi}_{\mathrm{r}}(t)$ are the rotor azimuth angle and rotational speed, respectively, $\phi(t)$ is the collective blade pitch angle, and the operator $\mathcal{A}(\cdot)$ denotes the aerodynamic load model, in which the blade-element momentum equations are solved at each time step. The turbulent wind field $\mathbf{u}_{\mathrm{w}}(\mathbf{r},t)$ is generated, for a specified mean wind speed, using TurbSim \cite{jonkman2014turbsim}. The wind turbine control adopts a conventional baseline control \cite{hansen2005control}, which uses the rotor speed as the feedback input and provides the generator torque and collective blade pitch commands. 

\if false
\begin{equation}
\begin{bmatrix}
T_{\mathrm{gen}}(t) \\
\phi(t)
\end{bmatrix}
=
\mathcal{C}
\left(
\dot{\psi}_{\mathrm{r}}(t)
\right),
\label{eq:baseline_control}
\end{equation}
where $T_{\mathrm{gen}}(t)$ is the generator torque, and the operator $\mathcal{C}(\cdot)$ denotes the turbine controller.
\fi

\subsubsection{Hydrodynamic loads}
The wave excitation load acting on the $i$-th body is simulated as
\begin{equation}
\mathbf{F}_{\mathrm{exc},i}(t)
=\sum_{k=1}^{N_{\omega}}\Re\left\{\mathbf{E}_i(j \omega_k)A(j \omega_k)e^{j \omega_k t}\right\}, \quad i=0,1,2,3,
\label{eq:excitation_force}
\end{equation}
where $\mathbf{E}_i(j\omega_k)$ is the complex frequency-domain excitation coefficient at frequency $\omega_k$, $A(j\omega_k)$ is the complex wave amplitude coefficient, and $N_{\omega}$ is the number of sampled frequencies. For a sea state specified by the significant wave height and peak wave period, the waves are modelled using the JONSWAP spectrum following IEC TS 62600-2 ED2 Annex C.2 (2019), with randomized phases assigned to the frequency components, to obtain $A(j\omega_k)$.

The radiation loads acting on all bodies, including mutual radiation interactions, can be jointly described by
\begin{equation}
\mathbf{F}_{\mathrm{rad}}(t)=-\mathbf{M}_{\infty}\ddot{\boldsymbol{\xi}}(t)
-\int_{-\infty}^{t}
\mathbf{K}_{\mathrm{rad}}(t-\tau)
\dot{\boldsymbol{\xi}}(\tau)
\,\mathrm{d}\tau,
\label{eq:radiation_force}
\end{equation}
where $\mathbf{M}_{\infty}$ is the infinite-frequency added mass matrix, and $\mathbf{K}_{\mathrm{rad}}(t)$ is the matrix-valued radiation retardation function; both are obtained from the frequency-dependent added mass $\mathbf{M}_{\mathrm{a}}(\omega)$ and radiation damping $\mathbf{R}_{\mathrm{a}}(\omega)$, as $\mathbf{M}_{\infty}=\lim_{\omega\to\infty}\mathbf{M}_{\mathrm{a}}(\omega)$ and $\mathbf{K}_{\mathrm{rad}}(t)=(2/\pi)\int_{0}^{\infty}\mathbf{R}_{\mathrm{a}}(\omega)\cos(\omega t)\,\mathrm{d}\omega$.

The hydrostatic restoring loads acting on the $i$-th body are described by
\begin{equation}
\mathbf{F}_{\mathrm{rest},i}(t)= \mathbf{F}_{\mathrm{g},i} + \mathbf{F}_{\mathrm{b},i} -\mathbf{K}_{\mathrm{hs},i}
\Delta \boldsymbol{\xi}_i(t), \quad i=0,1,2,3,
\label{eq:hydrostatic_force}
\end{equation}
where $\mathbf{F}_{\mathrm{g},i}$ and $\mathbf{F}_{\mathrm{b},i}$ are the gravitational and buoyancy loads, respectively, $\mathbf{K}_{\mathrm{hs},i}$ is the hydrostatic stiffness matrix, and $\Delta \boldsymbol{\xi}_i(t)$ denotes the pose deviation from static equilibrium.

Within the WEC-Sim framework, geometry meshes for all bodies are first established, and the boundary-element method software NEMOH \cite{kurnia2023nemoh} is then applied to calculate the hydrostatic and hydrodynamic parameters.

In addition, to compensate for the limitations of linear potential-flow theory, the viscous loads acting on the $i$-th body are modelled using drag terms as
\begin{equation}
\mathbf{F}_{\mathrm{vis},i}(t)=-\mathbf{D}_{\mathrm{vis},i}\left|\dot{\boldsymbol{\xi}}_i(t)\right| \odot \dot{\boldsymbol{\xi}}_i(t), \quad i=0,1,2,3,
\label{eq:viscous_force}
\end{equation}
where $\mathbf{D}_{\mathrm{vis},i}$ is the quadratic drag coefficient matrix, and $\odot$ denotes element-wise product. For the platform, the drag coefficients are calibrated values provided by MOST \cite{sirigu2022development}. For the WECs, only the viscous force in the heave direction, which is the dominant motion direction, is considered, and the coefficient is determined from the Morison equation \cite{morison1950force} as $D_{\mathrm{vis,z},i}=(1/2)\rho C_{\mathrm{d}} A_{\mathrm{cs}}$, $i=1,2,3$, where $\rho$ is the water density, $A_{\mathrm{cs}}$ is the WEC cross-sectional area, and $C_{\mathrm{d}}$ is the drag coefficient, assumed to be 1 \cite{fusco2014hierarchical} in this study. 

\subsubsection{Mooring loads}
The mooring system can be modelled as a quasi-static catenary system within MOST, where the dynamic mooring inertia is neglected, and the line tensions are determined from the instantaneous platform pose as
\begin{equation}
\mathbf{F}_{\mathrm{moor}}(t)=\mathcal{M}\left( \boldsymbol{\xi}_0(t) \right),
\label{eq:mooring_force}
\end{equation}
where the operator $\mathcal{M}(\cdot)$ represents the mooring load model, which is evaluated by solving the static catenary equations for each mooring line. This formulation captures the nonlinear restoring effect of the mooring system, while avoiding the computational cost of a fully dynamic mooring model.

\subsubsection{Mechanical friction forces}
The mechanical friction force associated with each PTO is modelled using a linear damping term, namely,
\begin{equation}
F_{\mathrm{fric},i}(t)=-R_{\mathrm{fric}}\dot{\zeta}_i(t),\quad i=1,2,3,
\end{equation}
where $\zeta_i(t)$ denotes the relative heave displacement between the WEC and the platform, and $R_{\mathrm{fric}}$ is the friction damping coefficient. Since friction depends on the detailed mechanical design, friction modelling is highly device-specific and difficult to anticipate without experimental testing \cite{lin2022fast}. Therefore, a representative value is adopted in this study, where $R_{\mathrm{fric}}$ is assumed to be one-tenth of the peak radiation damping of the WEC in the heave direction; the value is listed in Table \ref{tab_platform}. 

\subsection{PTO characteristics}
The focused control task is to determine the optimal PTO forces to be applied on the WECs. However, considering the PTO force limit $F_{\max}$ imposed by the generator capacity, whose assumed value is listed in Table \ref{tab_platform}, the commanded PTO force needs to be saturated as
\begin{equation}
F_{\mathrm{PTO},i}(t)=\mathrm{clip}\left(F_{\mathrm{PTO},i}^0(t), -F_{\max}, F_{\max}\right),\quad i=1,2,3,
\end{equation}
where $F_{\mathrm{PTO},i}^0(t)$ is the PTO force commanded by the WEC controller, $F_{\mathrm{PTO},i}(t)$ is the actual PTO force acting on the WEC, and $\mathrm{clip}(\cdot)$ denotes the clipping function such that 
\begin{equation}
\mathrm{clip}(F,-F_{\max},F_{\max})=\min\{\max\{F,-F_{\max}\},F_{\max}\}.
\end{equation}

The captured mechanical power can be calculated as
\begin{equation}
P_{\mathrm{mech}}(t)=-\sum_{i=1}^3 F_{\mathrm{PTO},i}(t) \dot{\zeta}_i(t).
\end{equation}
The electrical efficiency of the generator is another important factor to account for. In this study, a simple yet representative electrical loss model \cite{lin2024loss} is applied, where the loss power is assumed to vary quadratically with the generator force, so the electrical power output can be described as
\begin{equation}
P_{\mathrm{elec}}(t)=P_{\mathrm{mech}}(t)-\sum_{i=1}^3 \alpha \left(F_{\mathrm{PTO},i}(t)\right)^2,
\end{equation}
where $\alpha$, listed in Table \ref{tab_platform}, is the loss coefficient assumed to yield an efficiency of 96\%, a reasonable efficiency level for electrical generators, at a representative generator operating point of (0.6 m/s, 2000 kN), selected to represent the main operating range.

\subsection{Preliminary analysis}
\begin{figure}
    \centering
    \includegraphics[width=1\columnwidth]{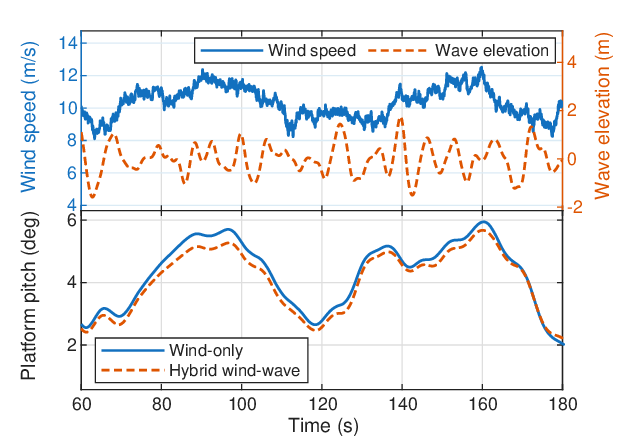}
    \caption{Motion trajectories of the wind-only and HWWE platforms, as well as the wind speed and wave elevation profiles.}
    \label{fig_wind_hybrid_traj}
\end{figure}

The HWWE platform with zero PTO control forces, i.e., with all WECs moving freely along the PTO guide, is first simulated and compared with the wind-only platform. The simulation sampling time is 0.02 s. The platform motion response is shown in Fig. \ref{fig_wind_hybrid_traj}, where the instantaneous turbulent wind speed and wave elevation are also illustrated, for reference. Under the wind load, the platform pitch motion, as the dominant rotational mode, exhibits a continuous bias. While no control forces are applied by the WEC PTOs, the HWWE platform exhibits slightly smaller pitch motion compared to the wind-only platform, which is partly due to the increase in the moment of inertia resulting from WEC integration; similar conclusions are also drawn in \cite{wang2025numerical}.

\section{Linear damping and reactive control}
Before introducing the RL controller, this section first defines the control performance metrics used throughout the study, and then develops the two benchmark control strategies, namely HOM and HET reactive control.

\subsection{Control evaluation metrics}
Two performance metrics are of primary concern in this study, when designing and evaluating control strategies. The first metric, $J_1$, reflects the wave energy production capability and is defined as the average electrical power output, namely,
\begin{equation}
J_1=\bar{P}=\frac{1}{T_{\mathrm{f}}}\int_0^{T_{\mathrm{f}}} P_{\mathrm{elec}}(t) \mathrm{d}t,
\label{eq:J1}
\end{equation}
where $\bar{P}$ is the average electrical power, and $T_{\mathrm{f}}$ is the terminal time of simulation evaluation. Note that, as is widely adopted by existing HWWE research \cite{sergiienko2025statistical}, the focus here is on wave energy capture rather than wind energy output, which is only indirectly affected by WEC motion; however, this effect will also be briefly examined later.

The second metric, $J_2$, characterizes the platform motion suppression effect and is defined as the root mean square (RMS) value of the platform pitch angle, namely,
\begin{equation}
J_2=\theta_{\mathrm{RMS}}
=\sqrt{\frac{1}{T_{\mathrm{f}}}
\int_0^{T_{\mathrm{f}}}\left(\theta^{\mathrm{y}}_0(t)\right)^2\mathrm{d}t},
\label{eq:J2}
\end{equation}
where $\theta_{\mathrm{RMS}}$ is the pitch RMS value. Note that different measures of platform pitch response have been adopted in previous studies, including the mean pitch drift \cite{wang2025numerical}, which reflects the shift in equilibrium position, and the pitch response amplitude operator \cite{han2024dynamic} or standard deviation \cite{zhu2022optimal}, which quantify the magnitude of dynamic oscillation. For simplicity and ease of comparison, this study uses the RMS value as a unified scalar metric. In particular, since $\theta_{\mathrm{RMS}}=\sqrt{\mathrm{mean}\left(\theta^{\mathrm{y}}_0\right)^2+\mathrm{std}\left(\theta^{\mathrm{y}}_0\right)^2}$, the RMS value can account for both the mean pitch drift and the dynamic pitch fluctuation.

\subsection{Homogeneous (HOM) damping and reactive control}
\begin{figure}
    \centering
    \includegraphics[width=1\columnwidth]{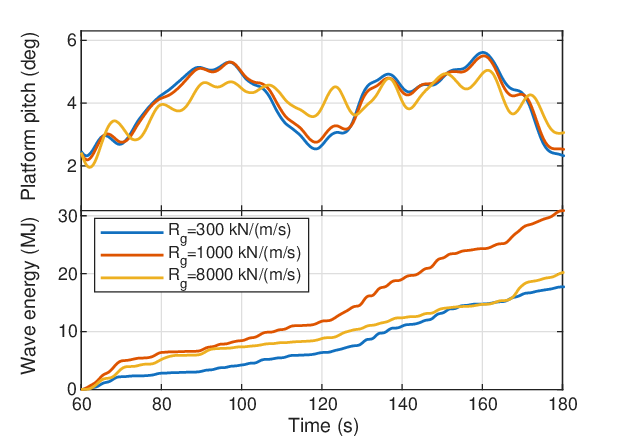}
    \caption{Damping control trajectories with different damping coefficients.}
    \label{fig_damping_traj}
\end{figure}

\begin{figure}
    \centering
    \includegraphics[width=1\columnwidth]{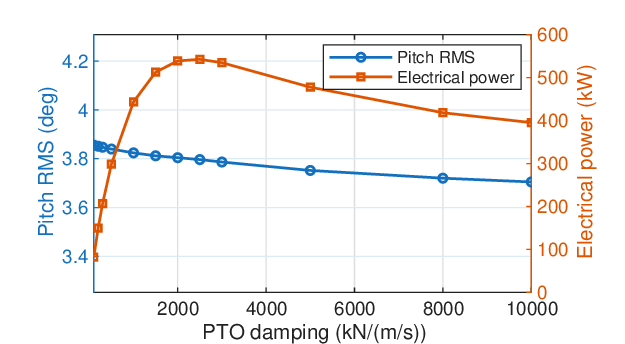}
    \caption{Impact of the damping coefficient.}
    \label{fig_damping}
\end{figure}

HOM control refers to the control scheme in which all three PTOs use identical control coefficients. In HOM damping control, the PTO forces are described by
\begin{equation}
F_{\mathrm{PTO},i}^0(t)
=-R_{\mathrm{g}}\dot{\zeta}_i(t),
\quad i=1,2,3,
\end{equation}
where $R_{\mathrm{g}}\geq 0$ is the uniform PTO damping coefficient. The platform motion and energy output trajectories, under different $R_{\mathrm{g}}$ values, are shown in Fig. \ref{fig_damping_traj}. When the damping is small, damping only slightly affects pitch motion, whereas a very large damping can reduce the pitch variation, while the change in mean drift is limited. On the other hand, a medium damping value yields the highest energy production, outperforming small- and large-damping control. These observations are further confirmed in Fig. \ref{fig_damping} where, regarding the pitch RMS value, the overall impact of damping is limited, while the energy production trend exhibits a typical single-peak pattern. It can be inferred that the energy and stability objectives are not totally antagonistic in HOM damping control. Without affecting pitch RMS too much, the optimal damping can be conveniently selected at the energy peak point, as in the case of standalone WEC control; similar approaches are also adopted in \cite{bayat2026multidisciplinary, cao2023wecs}. 

Alternatively, a stiffness term can be added to the PTO force, yielding HOM reactive control
\begin{equation}
F_{\mathrm{PTO},i}^0(t)
=-R_{\mathrm{g}}\dot{\zeta}_i(t)-K_{\mathrm{g}}\zeta_i(t),
\quad i=1,2,3,
\label{eq:hom_reactive_control}
\end{equation}
where $K_{\mathrm{g}}$ is the uniform PTO stiffness coefficient, and can be either positive or negative. From the perspective of the WEC, a positive stiffness increases the restoring force when the buoy deviates from its equilibrium position, whereas a negative stiffness reduces the effective restoring force. Notably, because point-absorber-type WECs usually have a natural period shorter than the dominant wave period, energy-maximizing control of a standalone WEC usually requires the PTO to provide a negative stiffness, exaggerating WEC motion.

\begin{figure}
    \centering
    \includegraphics[width=1\columnwidth]{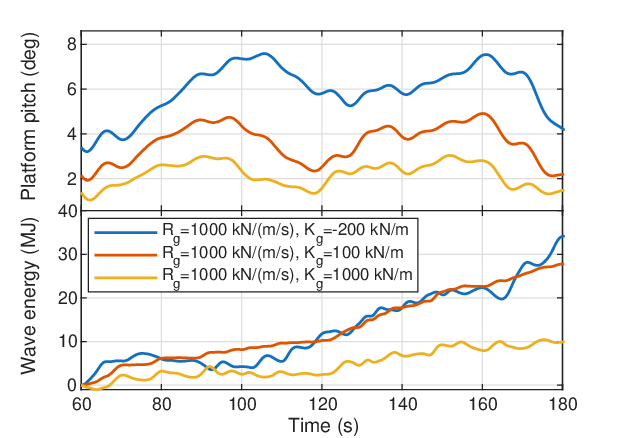}
    \caption{Reactive control trajectories with different stiffness coefficients.}
    \label{fig_reactive_traj}
\end{figure}

With this background, the control trajectories obtained with different PTO stiffness values are shown in Fig. \ref{fig_reactive_traj}. As expected, a smaller, more negative, stiffness yields higher energy generation. However, a smaller stiffness also leads to significantly larger platform pitch motion. This important phenomenon, consistent with previous analyses in \cite{hu2020optimal, si2021influence}, can be understood through a simple force analysis. When the platform pitches away from its initial position, the hydrostatic equilibrium positions of the WECs deviate from the PTO zero positions used in reactive control \eqref{eq:hom_reactive_control}. A positive PTO stiffness then generates a force that tends to restore the WECs toward those initial positions. The corresponding reaction force acting on the platform, in turn, produces a restoring moment that tends to drive the platform back toward its initial pitch angle. In other words, a positive stiffness provides a stabilizing effect on the platform. In contrast, a negative PTO stiffness tends to amplify the platform pitch deviation, exerting destabilizing effects.

\begin{figure}
    \centering
    \includegraphics[width=1\columnwidth]{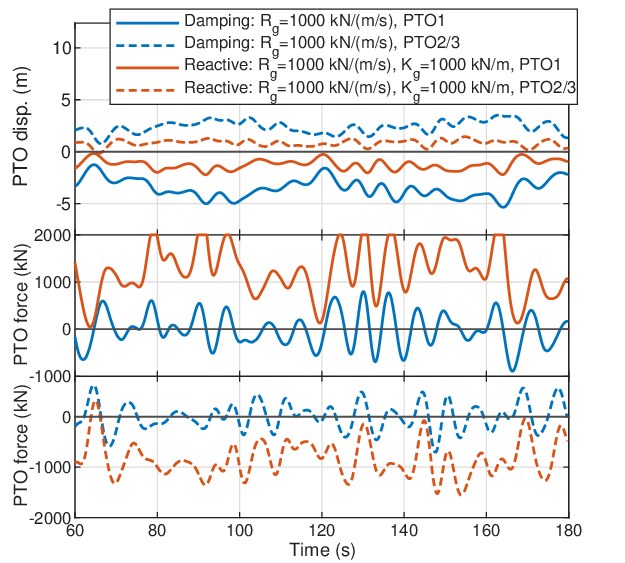}
    \caption{Control trajectories of damping and reactive control.}
    \label{fig_damping_reactive_traj}
\end{figure}

The impact of PTO stiffness is further illustrated in Fig. \ref{fig_damping_reactive_traj}, which compares the buoy displacement and PTO force under damping control and positive-stiffness reactive control. Reactive control maintains smaller WEC displacements by applying an additional restoring force, with a continuous, non-zero bias. Apparently, such force profiles are associated with larger reactive power exchange and higher electrical losses, and are therefore less favorable for energy production.

\begin{figure}
    \centering
    \includegraphics[width=1\columnwidth]{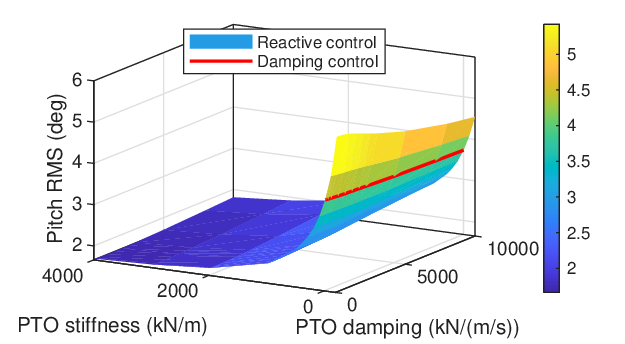}
    \includegraphics[width=1\columnwidth]{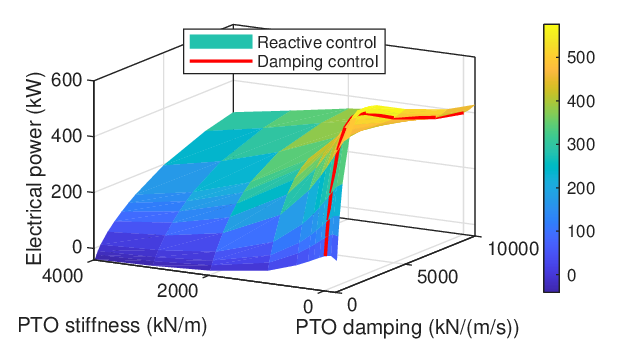}
    \caption{Parametric sweep result of HOM reactive control coefficients.}
    \label{fig_reactive}
\end{figure}

The overall impact of PTO coefficient values is shown in the parametric sweep result in Fig. \ref{fig_reactive}, where damping control, as a special case of reactive control with zero stiffness, is highlighted by the red lines. It is confirmed that, overall, decreasing the stiffness toward more negative values increases energy generation, but also amplifies platform pitch motion. Hence, in contrast to the damping control case, energy and stability appear to be conflicting objectives for reactive control, mainly due to the stiffness effect, as analyzed previously. A stiffness value smaller than the range shown in Fig. \ref{fig_reactive} will cause very significant platform pitch motion, and is therefore not considered in the study. 

\begin{figure}
    \centering
    \includegraphics[width=1\columnwidth]{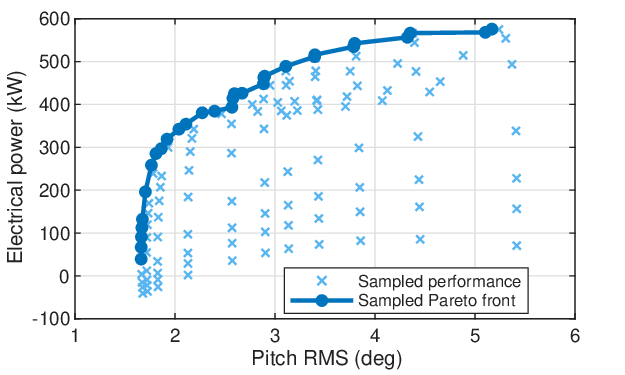}
    \caption{Performance samples of HOM reactive control, obtained from parametric sweep.}
    \label{fig_hom_scatter_pareto}
\end{figure}

The tradeoff between energy capture and platform motion suppression can be jointly described using a Pareto front, which denotes the performance boundary along which no objective can be further improved without degrading the other. The parametric sweep results, as well as the Pareto front defined by the swept samples, are shown in Fig. \ref{fig_hom_scatter_pareto}. Overall, HOM reactive control is restricted by a limited degree of freedom, with a linear control law and only two tunable parameters. Further control optimization would, ideally, push the Pareto front to the upper left. 

\subsection{Heterogeneous (HET) reactive control}
Hereafter, damping control is treated as a special case of reactive control, and both are uniformly referred to as reactive control. In contrast to HOM reactive control, HET reactive control means that different PTOs utilize different control coefficients. In this study, since the HWWE platform is oriented toward the predominant wind-wave direction (both assumed to be the x-axis direction), the two down-wave WECs, namely WECs 2 and 3, experience nearly identical operating conditions, whereas the up-wave WEC, namely WEC 1, operates under a different condition. Hence, two sets of reactive control coefficients can be applied for PTO 1 and PTO 2, 3, respectively, as
\begin{equation}
\begin{aligned}
F_{\mathrm{PTO},i}^0(t)
&=-R_{\mathrm{g},1}\dot{\zeta}_i(t)-K_{\mathrm{g},1}\zeta_i(t), \quad i=1 \\
F_{\mathrm{PTO},i}^0(t)
&=-R_{\mathrm{g},2}\dot{\zeta}_i(t)-K_{\mathrm{g},2}\zeta_i(t), \quad i=2,3,
\end{aligned}
\end{equation}
where $R_{\mathrm{g},1/2}$, $K_{\mathrm{g},1/2}$ are the PTO damping and stiffness values, respectively. A similar heterogenous control setting is also adopted in \cite{wang2026enhancing, li2022power}. Notably, HET reactive control introduces two additional tunable parameters, thereby offering greater flexibility to improve control performance. 

However, since HET control involves four degrees of freedom, the evaluation of control coefficients is not as straightforward as in HOM control, for which a simple 2-D parametric sweep can be applied. In particular, the evaluation of each control candidate requires one round of numerical simulation, so the resulting computational cost would be prohibitive if a 4-D parameter sweep is to be adopted. Moreover, even in the 2-D case, a parametric sweep cannot identify the exact Pareto front, since only a predefined grid of discrete coefficient values is evaluated. Hence, to identify the Pareto fronts for both HOM and HET reactive controllers, a multi-objective Bayesian optimization framework, based on the batch expected hypervolume improvement (qEHVI) algorithm \cite{daulton2020differentiable}, is adopted. The basic principle is to construct Gaussian process (GP) surrogate models of the objective functions, from limited simulation data, and use GP-predicted distributions to identify the most promising control candidates by maximizing the expected improvement in Pareto hypervolume. Detailed associated algorithmic procedures are outlined in the Appendix. 

\begin{figure}
    \centering
    \includegraphics[width=1\columnwidth]{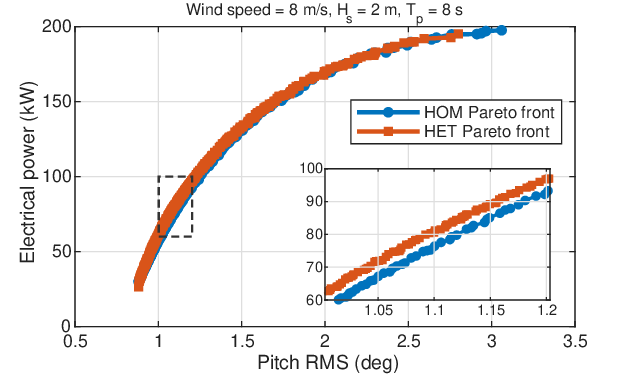}
    \includegraphics[width=1\columnwidth]{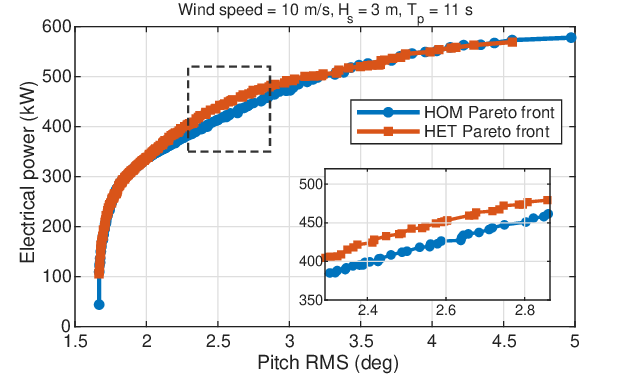}
    \includegraphics[width=1\columnwidth]{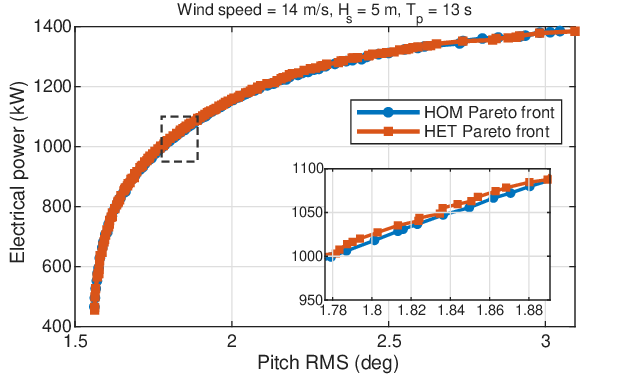}
    \caption{Pareto fronts of the HOM and HET reactive controllers, obtained using Bayesian optimization.}
    \label{fig_hom_het_qEHVI_pareto}
\end{figure}

The Pareto optimization results, for the three considered wind-wave conditions selected from Fig. \ref{fig_wind_wave_dist}, are shown in Fig. \ref{fig_hom_het_qEHVI_pareto}. It can be seen that HET control, with two additional degrees of freedom, improves the Pareto front beyond HOM control. However, this improvement is relatively limited and occurs mainly in the intermediate range of the two objectives. This result indicates that, within the linear control framework, introducing spatially heterogeneous control coefficients provides only moderate additional flexibility, and that both HOM and HET control remain rather constrained in handling the complex dynamics of the HWWE platform.

\section{Reinforcement-learning-based control}
To overcome the performance limitations of conventional control methods for the HWWE platform, this section proposes a reinforcement learning (RL)-based control framework, aiming at directly learning a real-time (nonlinear) control policy from interactions with the high-fidelity numerical model.
\subsection{Outline of PPO algorithm}
In RL, the control problem is formulated as a Markov decision process in which, at each control step, the controller observes the system state $\mathbf{s}_k$, generates an action $\mathbf{a}_k$, receives an immediate reward $r_k$, and arrives at a next state $\mathbf{s}_{k+1}$. The objective is to learn a control policy, namely, a mapping from state to action, that maximizes the expected cumulative reward.

Proximal policy optimization (PPO) employs an actor-critic architecture to achieve this objective \cite{schulman2017proximal}. The actor $\pi_{\boldsymbol{\theta}}(\mathbf{a}_k|\mathbf{s}_k)$ represents the control policy, whereas the critic $V_{\boldsymbol{\phi}}(\mathbf{s}_k)$ estimates the expected cumulative reward from state $\mathbf{s}_k$; here, $\boldsymbol{\theta}$ and $\boldsymbol{\phi}$ denote the parameters of the actor and critic neural networks, respectively. For continuous control action, the actor represents the action using a diagonal Gaussian distribution,
\begin{equation}
\mathbf{a}_k
\sim
\pi_{\boldsymbol{\theta}}(\cdot|\mathbf{s}_k)
=
\mathcal{N}
\left(
\boldsymbol{\mu}_{\boldsymbol{\theta}}(\mathbf{s}_k),
\operatorname{diag}
\left[
\boldsymbol{\sigma}_{\boldsymbol{\theta}}^2(\mathbf{s}_k)
\right]
\right),
\label{eq:ppo_policy}
\end{equation}
where $\boldsymbol{\mu}_{\boldsymbol{\theta}}$ and $\boldsymbol{\sigma}_{\boldsymbol{\theta}}$ are generated by the actor network.

During each episode with $N$ steps, the frozen old policy, $\pi_{\boldsymbol{\theta}_{\mathrm{old}}}$, interacts with the environment and collects a trajectory $\{\mathbf{s}_k,\mathbf{a}_k,r_k, \mathbf{s}_{k+1}\}_{k=0}^{N-1}$. The frozen old critic, $V_{\boldsymbol{\phi}_{\mathrm{old}}}$, is then used to evaluate how much better or worse each sampled action is, compared with the expected return at the corresponding state. This relative performance is quantified, using generalized advantage estimation (GAE), as
\begin{align}
\delta_k
&=
r_k+\gamma V_{\boldsymbol{\phi}_{\mathrm{old}}}(\mathbf{s}_{k+1})
-V_{\boldsymbol{\phi}_{\mathrm{old}}}(\mathbf{s}_k), \\
\hat{A}_k
&=
\delta_k+\gamma\lambda\hat{A}_{k+1},
\label{eq:gae}
\end{align}
where $\gamma$ is the discount factor for future rewards, $\delta_k$ is the temporal-difference error, $\lambda$ is the GAE coefficient, and $\hat{A}_k$ is the advantage estimate. A positive $\hat{A}_k$ indicates that the sampled action performs better than expected, and vice versa.

According to the policy gradient principle, the actor should increase the probability of actions with positive advantages and decrease that of actions with negative advantages. PPO performs this update using the probability ratio
\begin{equation}
\rho_k(\boldsymbol{\theta})
=
\frac{
\pi_{\boldsymbol{\theta}}(\mathbf{a}_k|\mathbf{s}_k)
}{
\pi_{\boldsymbol{\theta}_{\mathrm{old}}}
(\mathbf{a}_k|\mathbf{s}_k)
}
\label{eq:ppo_ratio}
\end{equation}
and maximizes the clipped surrogate objective
\begin{equation}
\begin{aligned}
&L^{\mathrm{clip}}(\boldsymbol{\theta})\\
&=
\frac{1}{N}
\sum_{k=0}^{N-1}
\min
\left\{
\rho_k(\boldsymbol{\theta})\hat{A}_k,\,
\operatorname{clip}
\left(
\rho_k(\boldsymbol{\theta}),1-\epsilon,1+\epsilon
\right)\hat{A}_k
\right\},
\label{eq:ppo_clip}
\end{aligned}
\end{equation}
where $\epsilon$ is the clipping coefficient. The clipping operation prevents excessively large changes between the updated and rollout policies, thereby improving training stability \cite{schulman2017proximal}. 

An entropy regularization term is further included to maintain exploration. The actor $\pi_{\boldsymbol{\theta}}$ is therefore updated by minimizing
\begin{equation}
\mathcal{L}_{\mathrm{actor}}=
-L^{\mathrm{clip}}(\boldsymbol{\theta})
-c_{\mathrm{ent}}
\frac{1}{N}
\sum_{k=0}^{N-1}
\mathcal{H}
\left[
\pi_{\boldsymbol{\theta}}(\cdot \mid \mathbf{s}_k)
\right],
\end{equation}
where $\mathcal{H}[\cdot]$ denotes the policy entropy and $c_{\mathrm{ent}}$ is the entropy coefficient. The entropy term prevents the action distribution from collapsing into an overly deterministic policy throughout training, sustaining exploration capability \cite{schulman2017proximal}.

Meanwhile, the critic $V_{\boldsymbol{\phi}}$ is trained to approximate the return associated with each state. Defining the return target as $V^*_k=\hat{A}_k+V_{\boldsymbol{\phi}_{\mathrm{old}}}(\mathbf{s}_k)$, the critic is updated by minimizing
\begin{equation}
\mathcal{L}_{\mathrm{critic}}
=
\frac{1}{N}
\sum_{k=0}^{N-1}
\left(
V_{\boldsymbol{\phi}}(\mathbf{s}_k)-V^*_k
\right)^2.
\label{eq:ppo_critic_loss}
\end{equation}

After actor and critic optimizations, the updated actor is used as the rollout policy for the next episode.

\subsection{RL problem formulation}
While RL algorithms have been well established through extensive studies, their practical application requires a dedicated formulation of the specific control task within the RL framework. For the HWWE system, the RL environment is directly provided by the numerical simulation model, without the need for model simplifications, as in \cite{zhu2022optimal}. A sampling period of $T_{\mathrm{s}}$=0.2 s is selected for the RL agent. 

The control action is defined straightforwardly as the (normalized) commanded forces for the three WECs, namely,
\begin{equation}
\mathbf{a}_k := \frac{1}{\mathcal{S}_{\mathrm{F}}}\left[F_{\mathrm{PTO},1,k}^0, F_{\mathrm{PTO},2,k}^0, F_{\mathrm{PTO},3,k}^0 \right]^\mathrm{T},
\end{equation}
where $\mathcal{S}_{\mathrm{F}}$ is the scaling factor for normalization, as listed in Table \ref{tab_RL}. The action remains constant over each control interval, according to the zero-order-hold convention. 

The selection of observation variables requires further consideration, since the selected variables should provide an approximate Markov representation of system dynamics that are relevant to the control objectives. It is well known that, in addition to WEC motion measurements, optimal WEC control, under irregular waves, generally relies on instantaneous and future information on the incident waves \cite{chen2024design,lin2025efficient}. Meanwhile, platform motion control requires direct platform motion measurements, and can further benefit from wind speed information and the blade pitch state. In this study, it is reasonably assumed that the available measurements include the wave elevations at the WEC locations, which can be obtained from up-wave buoy measurements and wave field reconstruction models \cite{belmont2006filters}, and the wind speed, which can be obtained from a nacelle-mounted anemometer. Accordingly, the state is defined as
\begin{equation}
\mathbf{s}_k
:=
\left[
\frac{\boldsymbol{\zeta}_k^{\mathrm{T}}}{\mathcal{S}_{\mathrm{\zeta}}},
\frac{\dot{\boldsymbol{\zeta}}_k^{\mathrm{T}}}{\mathcal{S}_{\mathrm{\dot{\zeta}}}},
\frac{\boldsymbol{\eta}_{k}^{\mathrm{T}}}{\mathcal{S}_{\mathrm{\eta}}},
\frac{\dot{\boldsymbol{\eta}}_{k}^{\mathrm{T}}}{\mathcal{S}_{\mathrm{\dot{\eta}}}},
\frac{\theta^{\mathrm{y}}_{0,k}}{\mathcal{S}_{\mathrm{\theta}}},
\frac{\dot{\theta}^{\mathrm{y}}_{0,k}}{\mathcal{S}_{\mathrm{\dot{\theta}}}},
\frac{\mathbf{u}_{\mathrm{w},k}^{\mathrm{T}}}{\mathcal{S}_{\mathrm{u}}},
\frac{\phi_k}{\mathcal{S}_{\mathrm{\phi}}}
\right]^{\mathrm{T}},
\label{eq:rl_state}
\end{equation}
which includes the positions of the three buoys $\boldsymbol{\zeta}_k=\left[\zeta_{1,k},\zeta_{2,k},\zeta_{3,k}\right]^{\mathrm{T}}$ and the corresponding velocities $\dot{\boldsymbol{\zeta}}_k$, the wave elevations $\boldsymbol{\eta}_{k}=\left[\eta_{1,k}, \eta_{2,k}, \eta_{3,k} \right]^{\mathrm{T}}$ and their time derivatives $\dot{\boldsymbol{\eta}}_{k}$, the platform pitch angle $\theta^{\mathrm{y}}_{0,k}$ and angular velocity $\dot{\theta}^{\mathrm{y}}_{0,k}$, the wind velocity at the turbine hub $\mathbf{u}_{\mathrm{w},k}$, and the collective blade pitch angle $\phi_k$; the associated scaling factors $\mathcal{S}_{\mathrm{\zeta}}$, $\mathcal{S}_{\mathrm{\dot{\zeta}}}$, $\mathcal{S}_{\mathrm{\eta}}$, $\mathcal{S}_{\mathrm{\dot{\eta}}}$, $\mathcal{S}_{\mathrm{\theta}}$, $\mathcal{S}_{\mathrm{\dot{\theta}}}$, $\mathcal{S}_{\mathrm{u}}$, and $\mathcal{S}_{\mathrm{\phi}}$ are listed in Table \ref{tab_RL}. Specifically, the inclusion of the time derivatives of wave elevation is to incorporate certain wave dynamics for the RL agent \cite{chen2024design}. Also note that wave excitation forces are (reasonably) not assumed available, since estimation of excitation forces generally requires a dynamic model of the system \cite{lin2024sensitivity}. For the considered HWWE system, model-based excitation force estimation is difficult, due to the complex dynamics.

Meanwhile, the RL reward function is designed as the weighted sum of two objectives: energy production and platform motion suppression, respectively:
\begin{equation}
\begin{aligned}
r_k :&= \beta r_{\mathrm{e},k} + (1-\beta) r_{\mathrm{s},k}, \\
r_{\mathrm{e},k}&=\frac{1}{\mathcal{S}_{\mathrm{F}}\mathcal{S}_{\mathrm{\zeta}}}
\sum_{i=1}^3 \left[ -F_{\mathrm{PTO},i,k} \left( \zeta_{i,k+1}-\zeta_{i,k} \right)
- \alpha T_{\mathrm{s}}F_{\mathrm{PTO},i,k}^2 \right], \\
r_{\mathrm{s},k}&=-\frac{1}{\mathcal{S}_{\mathrm{\theta}}^2}\frac{T_{\mathrm{s}}}{2}\left( \left(\theta^{\mathrm{y}}_{0,k}\right)^2+\left(\theta^{\mathrm{y}}_{0,k+1}\right)^2 \right),
\end{aligned}
\end{equation}
where $\beta$ is the weighting factor, and $r_{\mathrm{e},k}$ and $r_{\mathrm{s},k}$ are discrete-time calculations of one-step electrical energy output and the integral of pitch motion squared, consistent with \eqref{eq:J1} and \eqref{eq:J2}, respectively. The weighting factor $\beta$ is a key control parameter that reflects the designer's preference about control performance, and will guide the RL agent toward different tradeoffs between the conflicting objectives of energy capture and platform stability.

\subsection{Training framework}
Following the RL formulation, a RL training framework is designed for the HWWE system, in which the PPO algorithm is implemented in PyTorch and interacts with the MOST simulation environment in MATLAB. Each training episode consists of a time-domain simulation lasting 20 peak wave periods. During the simulation, the actor obtained from the previous episode generates the PTO control actions at each control step. Note that, during the training process, the actions are sampled from the Gaussian policy, parameterized by the actor network to enable exploration. In the implementation of a trained actor, however, only the mean actions are used.

Different random seeds are adopted across episodes to generate distinct (randomized) time-domain wind and wave profiles. Throughout each simulation, the trajectories of state, action, and reward are collected. After the simulation, the discounted returns and GAEs are calculated from the trajectory and then used to update the actor and critic networks.

Both the actor and critic are fully connected neural networks with two hidden layers, each containing 128 neurons and using the rectified linear unit (ReLU) activation function. The actor contains two output branches for the mean and standard deviation of the Gaussian policy, respectively. A hyperbolic tangent (tanh) activation function is applied to the mean output to constrain the PTO force within the prescribed limit, while a sigmoid function is applied to the standard deviation output to ensure positiveness and boundedness.

After each episode, the actor and critic parameters are updated separately, using the Adam optimizer \cite{kingma2014adam} and full-batch gradient-based optimization over all collected trajectory samples. The optimization is repeated for predefined numbers of epochs, using prescribed learning rates. The updated actor is then used to generate the control actions in the subsequent episode. The key algorithm parameters are summarized in Table \ref{tab_RL}, which are selected through preliminary sensitivity analysis by considering both training performance and computational efficiency.

\begin{table}
\centering
\caption{RL training and normalization parameters.}
\label{tab_RL}
\begin{tabular}{lll}
\hline
Category & Parameter & Value \\
\hline
Algorithm & Number of episodes & 100 \\
 & Actor learning rate & 5e-4 \\
 & Critic learning rate & 5e-4 \\
 & Actor number of epoches & 10 \\
 & Critic number of epoches & 20 \\
 & Discount factor $\gamma$ & 0.99 \\
 & GAE coefficient $\lambda$ & 0.95 \\
 & Clip coefficient $\epsilon$ & 0.2 \\
 & Entropy coefficient $c_{\mathrm{ent}}$ & 0.01 \\
\hline
Scaling & PTO force $\mathcal{S}_{\mathrm{F}}=F_{\max}$ & 2000 kN \\
        & WEC position $\mathcal{S}_{\mathrm{\zeta}}$ & 3 m \\
        & WEC velocity $\mathcal{S}_{\mathrm{\dot{\zeta}}}$ & 3 m/s \\
        & Wave elevation $\mathcal{S}_{\mathrm{\eta}}$ & 5 m \\
        & Wave elevation derivative $\mathcal{S}_{\mathrm{\dot{\eta}}}$ & 5 m/s \\
        & Pitch angle $\mathcal{S}_{\mathrm{\theta}}$ & 0.05 rad \\
        & Pitch angular velocity $\mathcal{S}_{\mathrm{\dot{\theta}}}$ & 0.01 rad/s \\
        & Wind speed $\mathcal{S}_{\mathrm{u}}$ & 20 m/s \\
        & Blade pitch angle $\mathcal{S}_{\mathrm{\phi}}$ & 30 deg \\
\hline
\end{tabular}
\end{table}

\section{Validation results}
\subsection{Traning process}
\begin{figure}
    \centering
    \includegraphics[width=1\columnwidth]{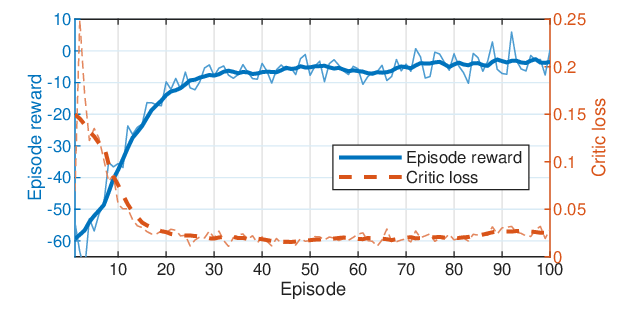}
    \caption{Example of the RL training process. The thin lines gives the true value, while the thick lines represent the moving average.}
    \label{fig_rl_training}
\end{figure}

A typical RL training process, with weighting factor $\beta=0.9$ and under the (WS, Hs, Tp)=(10 m/s, 3 m, 11 s) condition, is shown in Fig. \ref{fig_rl_training}. As the control policy is iteratively updated, the total episode reward increases gradually and then fluctuates around a relatively stable level. The fluctuation is mainly due to the randomized realization of wind and wave profiles. Meanwhile, the critic loss, which represents the value function approximation error, as defined in \eqref{eq:ppo_critic_loss}, remains relatively large during the early training stage, and then decreases as the actor approaches convergence, indicating that the critic provides an increasingly accurate representation of the value function. Overall, the learning process exhibits relatively stable convergence, which is partly attributed to the episode-based training procedure, in which the actor and critic are updated using full-batch gradient-based optimization over the complete trajectory.

\begin{figure}
    \centering
    \includegraphics[width=1\columnwidth]{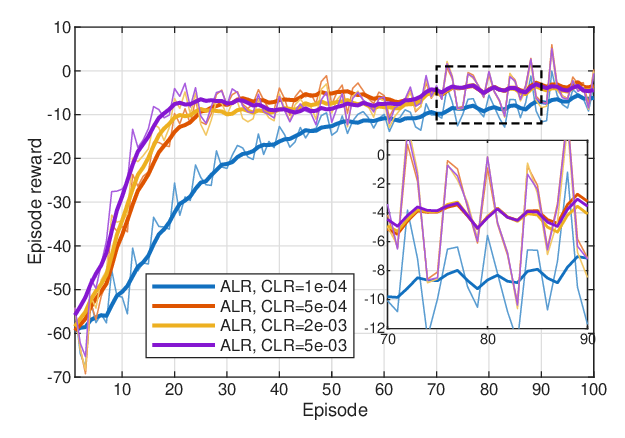}
    \caption{Impact of actor learning rate (ALR) and critic learning rate (CLR) on RL training performance.}
    \label{fig_rl_training_LRScan}
\end{figure}

A sensitivity analysis of RL hyperparameters is further shown in Fig. \ref{fig_rl_training_LRScan}, where the impact of the actor and critic learning rates is examined. For simplicity, the same learning rate is used for both networks. In general, an excessively small learning rate results in slow convergence, whereas an excessively large learning rate slightly degrades the final performance, due to less stable network updates. Accordingly, a learning rate of 5e-4 is selected in this study. 

\begin{figure}
    \centering
    \includegraphics[width=1\columnwidth]{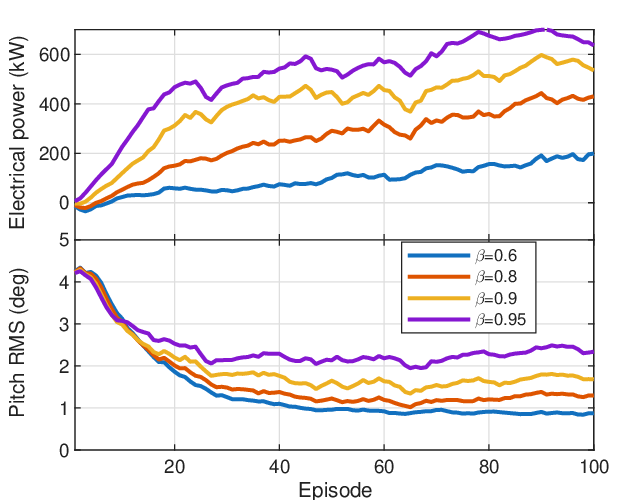}
    \caption{Evaluated electrical power and pitch RMS values of the actors obtained through the RL training process, under different weight factors $\beta$.}
    \label{fig_rl_training_eval}
\end{figure}

The actor obtained after each training episode is evaluated in a separate simulation, using wind and wave profiles generated with a random seed different from those used for training. Four values of the weighting factor $\beta$ are considered, with the corresponding evaluation results shown in Fig. \ref{fig_rl_training_eval}. As the training progresses, the RL agent gradually increases energy capture, and reduces platform pitch motion at the same time, consistent with the reward evolution shown in Fig. \ref{fig_rl_training}. A larger $\beta$ places greater emphasis on energy generation and therefore leads to a policy with higher energy output, but also larger platform pitch motion. These results again demonstrate the tradeoff between energy and stability objectives, and confirm that $\beta$ serves as a tuning parameter reflecting the preferred balance between the two objectives.

\subsection{Evaluation results}
\begin{figure}
    \centering
    \includegraphics[width=1\columnwidth]{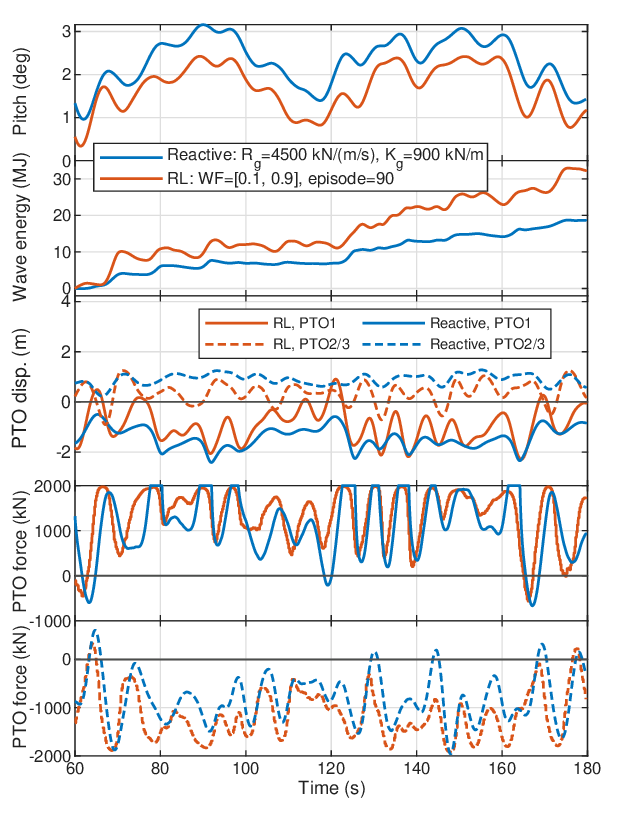}
    \caption{Example of control trajectories of HOM reactive control and RL control.}
    \label{fig_reactive_RL_traj}
\end{figure}

A trained actor, selected from the training process shown in Fig. \ref{fig_rl_training}, is compared with a HOM reactive controller selected from the Pareto front in Fig. \ref{fig_hom_het_qEHVI_pareto}, with parameters $[R_{\mathrm{g}}, K_{\mathrm{g}}]$=[4500 kN/(m/s), 900 kN/m] and performance $[\theta_{\mathrm{RMS}}, \bar{P}]\approx$[2.3 deg, 385 kW]; the control trajecories are shown in Fig. \ref{fig_reactive_RL_traj}. The RL controller simultaneously reduces platform pitch motion and increases electrical energy generation. Note that, while RL control results in smaller buoy displacement and greater PTO force efforts, with increased electrical losses, net electrical energy is still improved. Moreover, the PTO force constraint is strictly satisfied through the bounded output of the actor network. This comparison provides a straightforward demonstration that the RL controller can improve both objectives simultaneously, beyond the performance limit of HOM reactive control.

\begin{figure}
    \centering
    \includegraphics[width=1\columnwidth]{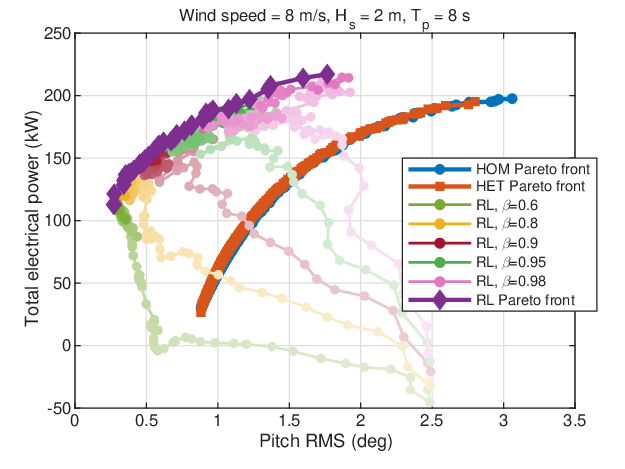}
    \includegraphics[width=1\columnwidth]{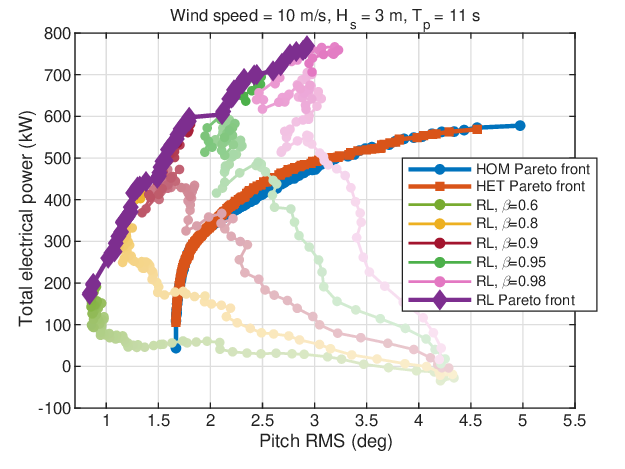}
    \includegraphics[width=1\columnwidth]{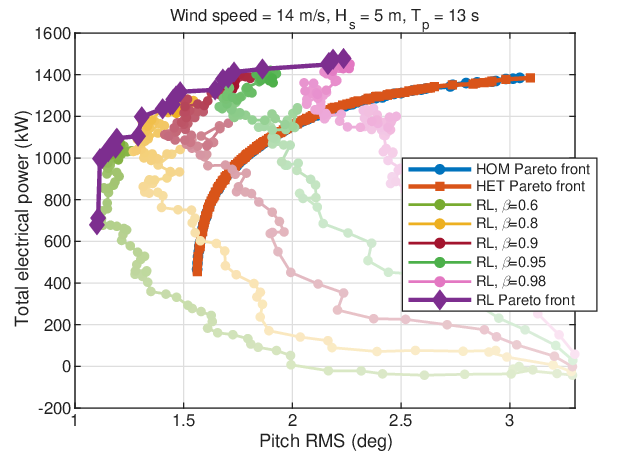}
    \caption{Pareto fronts for RL control, compared with HOM and HET reactive control. For RL control, the performance of each actor during the iteration process is displayed, with colors from light to dark corresponding to the episode order.}
    \label{fig_hom_het_rl_pareto}
\end{figure}

To comprehensively validate the effectiveness of RL control, in a Pareto sense, all three wind-wave conditions selected from Fig. \ref{fig_wind_wave_dist} are considered, together with different values of the weighting factor $\beta$. For each wind-wave condition, the Pareto fronts obtained using HOM reactive control, HET reactive control, and RL control are jointly compared in Fig.~\ref{fig_hom_het_rl_pareto}. 

The performance improvement brought by RL is impressive. First, the control preference posed by the weight factor $\beta$ can be observed again, which directs the performance towards different regions. The performance points, obtained with different $\beta$, jointly form the RL control Pareto front. It is clear that RL achieves significantly improved Pareto values, compared with the HOM and HET reactive controllers, under all the considered wind-wave conditions; in other words, RL control offers a superior performance tradeoff between energy capture and platform motion suppression for control designers. For example, for the second wind-wave condition in Fig. \ref{fig_hom_het_rl_pareto}, for the same level of platform pitch motion (2.3 deg), RL control is capable of generating $\sim$75\% more wave energy, compared to HOM and HET reactive control; conversely, for the same lavel of energy generation (385 kW), RL control can reduce the platform pitch RMS value by nearly 50\%. Such improved performance mainly arises from the nonlinear, real-time control law represented by the RL policy, and from the direct learning through interactions with the numerical model. Also, it should be acknowledged that RL also benefits from the inclusion of more observation signals than reactive control. 

In addition to wave energy capture and platform pitch motion, which are two primary control objectives considered in most HWWE studies, other relevant performance metrics are also evaluated, including wind power generation and the remaining platform motion modes. The same HOM reactive and RL controllers examined in Fig. \ref{fig_reactive_RL_traj} are compared, with the results summarized in Table \ref{tab_other_metrics}. By influencing platform motion, the RL controller yields a slightly higher wind power output than reactive control, although this improvement is rather limited compared with the increase in wave energy capture. RL control also slightly reduces platform heave while increasing roll motion. This increase may arise because the RL formulation does not explicitly impose an identical (symmetric) policy for WECs 2 and 3; consequently, the learned controller may apply different PTO forces to the two WECs, introducing asymmetric loading and thereby increasing the roll response. Changes are also observed in surge, sway, and yaw; however, because the WECs operate primarily in the vertical direction, these modes are less directly affected by WEC control, and the corresponding differences remain limited.

\begin{table}
\centering
\caption{Overall performance comparison between the selected reactive and RL controllers.}
\label{tab_other_metrics}
\begin{tabular}{lll}
\hline
Performance metric & Reactive & RL \\
\hline
Wind power avg. (MW) & 11.623 & 11.671 \\
Wave power avg. (MW) & 0.386 & 0.598 \\
Surge RMS (m) & 19.338 & 19.567 \\
Sway RMS (m) & 0.895 & 0.793 \\
Heave RMS (m) & 0.307 & 0.296 \\
Roll RMS (deg) & 0.227 & 0.319 \\
Pitch RMS (deg) & 2.309 & 1.796 \\
Yaw RMS (deg) & 1.686 & 1.762 \\
\hline
\end{tabular}
\end{table}

\section{Conclusion}
This study proposes an RL-based control framework for WECs in HWWE systems, with the objective of increasing wave energy production while suppressing the motion of the floating platform. First, a numerical simulation model is established using MOST, incorporating coupled aero-hydro-structural-mooring dynamics, together with realistic PTO characteristics, such as force limits and electrical losses. As benchmark strategies, HOM linear damping and reactive control are first investigated. It is shown that PTO damping mainly affects energy capture while having a limited influence on platform pitch RMS. PTO stiffness, in contrast, has a pronounced effect on both objectives. Specifically, small or negative stiffness can enhance energy capture but is detrimental to platform stability, whereas large, positive stiffness produces the opposite effect, revealing a strong tradeoff between energy generation and motion suppression objectives. Heterogeneous (HET) reactive control is further introduced as an extension of HOM control with greater control flexibility. The Pareto fronts for HOM and HET reactive control are identified using Bayesian optimization. The results show that HOM reactive control is constrained by a limited performance tradeoff, while the additional flexibility of HET control yields rather limited improvement.

Importantly, an RL control framework, based on the PPO algorithm, is then developed. The control input includes the motion states of the WECs and the FOWT, together with relevant wind and wave variables, while the reward function is formulated as a weighted combination of wave energy capture and platform motion suppression. The RL agent learns the control policy through direct interaction with the numerical model. With appropriate tuning of the training parameters, stable convergence of the RL algorithm is achieved, and the weighting factor in the reward function effectively guides the agent to learn control policies with different preferences between the two objectives. Comprehensive testing under different wind-wave conditions demonstrates that RL control achieves substantially improved Pareto fronts compared with both HOM and HET reactive control. For example, at the same level of platform oscillation, RL control can generate more than 75\% additional wave energy; conversely, for the same wave energy generation, RL control can reduce the platform pitch RMS by nearly 50\%. In addition, RL control only has limited impacts on wind power generation and other platform motion modes. Overall, RL control proves to be a powerful approach for extending the attainable performance envelope of HWWE systems. 

\section{Acknowledgements}
This wors was supported by the Sustainable Energy Authority of Ireland through the RDD Energise Fellowship under Grant No. 25/RDDF/793.

\appendix
\section{Multi-objective Bayesian optimization}
The aim of the optimization is to identify the Pareto front, with respect to wave power maximization and platform pitch RMS minimization, using a limited number of computationally demanding, yet parallelizable, simulations. The optimization is performed using the qEHVI method \cite{daulton2020differentiable}. This framework is applied to both the HOM and HET reactive control.

Take the HET reactive control as an example, the optimization variables are defined as
\begin{equation}
\mathbf{x}
=\left[
\log_{10}\left(R_{\mathrm{g},1}\right),
K_{\mathrm{g},1},
\log_{10}\left(R_{\mathrm{g},2}\right),
K_{\mathrm{g},2}
\right]^{\mathrm T},
\end{equation}
and the two maximization objectives are
\begin{equation}
f_1(\mathbf{x})=-\theta_{\mathrm{RMS}}(\mathbf{x}),\quad 
f_2(\mathbf{x})=\bar{P}(\mathbf{x}).
\end{equation}
The design variables and objective values are further normalized to the interval $[0,1]$, according to the prescribed bounds.

\paragraph{Step 1: Initial sampling and simulation evaluation.}
An initial set of $N_0$ design points, $\left\{ \mathbf{x}^{(1)},\ldots,\mathbf{x}^{(N_0)}\right\}$, is first generated over the parameter domain, using Latin hypercube sampling (LHS). Each point is evaluated using the numerical model, forming the initial dataset 
\begin{equation}
\mathcal{D}_0
=\left\{
\mathbf{x}^{(i)},
\mathbf{f}(\mathbf{x}^{(i)})
\right\}_{i=1}^{N_0},
\end{equation}
where $\mathbf{f}=[f_1,f_2]^\mathrm{T}$.

\paragraph{Step 2: Gaussian-process surrogate modelling.}
In the $n$-th iteration of Bayesian optimization, suppose that $N_n$ design points have been evaluated, with the available dataset
\begin{equation}
\mathcal{D}_n
=
\left\{
\mathbf{x}^{(i)},
\mathbf{f}(\mathbf{x}^{(i)})
\right\}_{i=1}^{N_n}.
\end{equation}
The two objective functions are modelled, independently, as random functions using Gaussian processes (GPs):
\begin{equation}
f_m(\cdot)
\sim
\mathcal{GP}
\left(
m_m(\cdot),
k_m(\cdot,\cdot)
\right),
\qquad m=1,2,
\label{eq:gp_prior_qehvi}
\end{equation}
where
$m_m(\mathbf{x})=\mathbb{E}[f_m(\mathbf{x})]$
is the mean function, and
$k_m(\mathbf{x},\mathbf{x}')
=
\operatorname{Cov}
\left[
f_m(\mathbf{x}),
f_m(\mathbf{x}')
\right]$
is the covariance function between the objective values at two given design points $\mathbf{x}$ and $\mathbf{x}'$. A constant mean function and an automatic relevance determination squared-exponential (ARD-SE) covariance kernel function are adopted
\begin{equation}
\begin{aligned}
m_m(\mathbf{x})
&=
\beta_m,\\
k_m(\mathbf{x},\mathbf{x}')
&=
\sigma_{f,m}^{2}
\exp
\left[
-\frac{1}{2}
\sum_{d=1}^{4}
\frac{
\left(x_d-x_d'\right)^2
}{
\ell_{m,d}^{2}
}
\right],
\qquad m=1,2,
\end{aligned}
\label{eq:gp_kernel_qehvi}
\end{equation}
where $\beta_m$ is the constant mean, $\sigma_{f,m}^{2}$ is the kernel variance, $x_d$ and $x_d'$ denote the $d$th elements of $\mathbf{x}$ and $\mathbf{x}'$, respectively, and $\ell_{m,d}$ is the characteristic length scale of the $d$th design variable for objective $m$. Hence, the covariance between two objective values decreases, as the corresponding design points become farther apart.

For objective $m$, the evaluated $N_n$ design points and the corresponding objective values form the following matrices
\begin{equation}
\mathbf{X}_n
=
\begin{bmatrix}
\left(\mathbf{x}^{(1)}\right)^{\mathrm{T}}\\
\vdots\\
\left(\mathbf{x}^{(N_n)}\right)^{\mathrm{T}}
\end{bmatrix},\quad
\mathbf{y}_{m,n}
=
\begin{bmatrix}
f_m(\mathbf{x}^{(1)})\\
\vdots\\
f_m(\mathbf{x}^{(N_n)})
\end{bmatrix}.
\label{eq:gp_training_data}
\end{equation}
Consider any set of $N_*$ unevaluated design points
\begin{equation}
\mathbf{X}_{*}
=
\begin{bmatrix}
\left(\mathbf{x}_{*}^{(1)}\right)^{\mathrm{T}}\\
\vdots\\
\left(\mathbf{x}_{*}^{(N_*)}\right)^{\mathrm{T}}
\end{bmatrix}.
\end{equation}
Conditioning the GP model on the evaluated dataset $\mathcal{D}_n$ gives the joint posterior distribution
\begin{equation}
\mathbf{f}_m(\mathbf{X}_{*})
\mid
\mathcal{D}_n
\sim
\mathcal{N}
\left(
\boldsymbol{\mu}_{m,n}(\mathbf{X}_{*}),
\boldsymbol{\Sigma}_{m,n}(\mathbf{X}_{*})
\right),
\qquad m=1,2,
\label{eq:gp_posterior_qehvi}
\end{equation}
where
$\mathbf{f}_m(\mathbf{X}_{*})=\left[f_m(\mathbf{x}_{*}^{(1)}),...,f_m(\mathbf{x}_{*}^{(N_*)})\right]^{\mathrm{T}}$, and the posterior mean and covariance can be calculated as
\begin{equation}
\begin{aligned}
\boldsymbol{\mu}_{m,n}(\mathbf{X}_{*})
=&
\beta_m\mathbf{1}_{N_*}+\mathbf{K}_m(\mathbf{X}_{*},\mathbf{X}_n)\mathbf{C}_{m,n}^{-1}
\left(
\mathbf{y}_{m,n}-\beta_m\mathbf{1}_{N_n}
\right),\\
\boldsymbol{\Sigma}_{m,n}(\mathbf{X}_{*})
=&
\mathbf{K}_m(\mathbf{X}_{*},\mathbf{X}_{*})
-
\mathbf{K}_m(\mathbf{X}_{*},\mathbf{X}_n)
\mathbf{C}_{m,n}^{-1}
\mathbf{K}_m(\mathbf{X}_n,\mathbf{X}_{*})\\
\mathbf{C}_{m,n}
=&
\mathbf{K}_m(\mathbf{X}_n,\mathbf{X}_n)
+
\sigma_{\epsilon,m}^{2}\mathbf{I}_{N_n},
\end{aligned}
\label{eq:gp_posterior_mean_cov}
\end{equation}
where $\mathbf{K}_m(\mathbf{X}_*,\mathbf{X}_n)$ denotes the kernel matrix between the design points contained in $\mathbf{X}_*$ and $\mathbf{X}_n$, $\mathbf{1}_{N}$ is an $N$-dimensional vector of ones, $\mathbf{I}_{N_n}$ is the $N_n$-dimensional identity matrix, and $\sigma_{\epsilon,m}^{2}$ is a small number, introduced to improve numerical conditions. The posterior mean represents the predicted objective values, while the posterior covariance quantifies both the uncertainty of each prediction and the statistical correlation among different unevaluated designs.

\paragraph{Step 3: Construction of the current Pareto front.}
For the available dataset $\mathcal{D}_n$, the current approximation of Pareto front is defined as
\begin{equation}
\mathcal{P}_n
=
\left\{
\mathbf{f}(\mathbf{x}^{(i)}):
\nexists\,j\ \mathrm{such\ that}\
\mathbf{f}(\mathbf{x}^{(j)})\succ
\mathbf{f}(\mathbf{x}^{(i)})
\right\}.
\end{equation}
where $\mathbf{a}\succ\mathbf{b}$ denotes that $a_m\ge b_m$ for all objectives and $a_m>b_m$ for at least one objective.

The convergence and coverage of the current Pareto front are quantified by the hypervolume. Given a fixed reference point $\mathbf{r}$, the hypervolume is denoted as
\begin{equation}
\operatorname{HV}
\left(
\mathcal{P}_n;\mathbf{r}
\right),
\end{equation}
which, for the present two-objective problem, is the area of the objective space dominated by the Pareto front and bounded by a reference point. In implementation, the reference point is manually selected for each wind-wave condition, to represent a poor performance across all objectives.

\begin{table}
\centering
\caption{Bayesian optimization parameters.}
\label{tab_eqhvi_params}
\begin{tabular}{lll}
\hline
Parameter & HOM & HET \\
\hline
Damping range
& \multicolumn{2}{l}{[100,10000] kN/(m/s)} \\
Stiffness range
& \multicolumn{2}{l}{[-200,4000] kN/m} \\
Initial LHS samples & 40 & 100 \\
Number of iterations & 10 & 25 \\
Candidate pool size & 5000 & 15000 \\
Shortlist size & 150 & 300 \\
qEHVI MC samples, $N_{\mathrm{MC}}$ & 512 & 512 \\
Batch size, $q$ & 20 & 40 \\
\hline
\end{tabular}
\end{table}

\paragraph{Step 4: Selection of the next simulation batch.}
For an unevaluated candidate batch $\mathbf{X}_q$, the expected hypervolume improvement can be approximately calculated using Monte Carlo (MC) sampling from the joint GP posterior
\begin{equation}
\begin{aligned}
&\widehat{\alpha}_{q\mathrm{EHVI}}
\left(
\mathbf{X}_q
\right)\\
&=
\frac{1}{N_{\mathrm{MC}}}
\sum_{i=1}^{N_{\mathrm{MC}}}
\left[
\operatorname{HV}
\left(
\mathcal{P}_n
\cup
\mathbf{f}^{(i)}(\mathbf{X}_q);
\mathbf{r}
\right)
-
\operatorname{HV}
\left(
\mathcal{P}_n;\mathbf{r}
\right)
\right],
\end{aligned}
\label{eq:qehvi_mc}
\end{equation}
where $\mathbf{f}^{(i)}(\mathbf{X}_q)$ is the $i$th joint posterior sample, and $N_{\mathrm{MC}}$ is the number of samples. The most promising batch $\mathbf{X}_q^{*}$ is selected such that
\begin{equation}
\mathbf{X}_q^{*}
\approx \arg\max_{\mathbf{X}_q}
\widehat{\alpha}_{q\mathrm{EHVI}}
\left(
\mathbf{X}_q
\right)
\end{equation}
which, in implementation, is approximately maximized using a greedy approach based on \cite{daulton2020differentiable,wilson2018maximizing}. Specifically, a large candidate pool for $\mathbf{X}_q$ is first generated using LHS, and candidates close to previously evaluated design points are removed. A single-point evaluation of EHVI is then performed for each candidate alone, and the candidates with the highest EHVI values form a shortlist. Within the shortlist, the batch is constructed sequentially by selecting one candidate at each step, according to the greedy qEHVI approach in \cite{daulton2020differentiable, wilson2018maximizing}.

\paragraph{Step 5: Simulation evaluation and update.}
The selected batch of design parameters $\mathbf{X}_q^*$ are evaluated in parallel using the numerical simulation model. The new simulation results are then appended to the dataset as
\begin{equation}
\mathcal{D}_{n+1}
=
\mathcal{D}_n
\cup
\left\{
\mathbf{X}_q^{*},
\mathbf{f}(\mathbf{X}_q^{*})
\right\},
\end{equation}
and the above procedures are repeated iteratively for a prescribed number of times. The overall algorithmic parameters are listed in Table \ref{tab_eqhvi_params}. 

\bibliographystyle{IEEEtran}
\bibliography{references}
\balance


\end{document}